\documentclass[sigconf]{acmart}
\usepackage{graphicx}

\usepackage{bm}
\usepackage{setspace}
\usepackage{amsmath,amsfonts,amsthm}
\usepackage{booktabs}

\usepackage{textcomp}
\usepackage{xcolor}
\usepackage[font=footnotesize]{subfig}
\usepackage{multirow}
\usepackage{float}

\setcopyright{acmcopyright}
\AtBeginDocument{%
  \providecommand\BibTeX{{%
    \normalfont B\kern-0.5em{\scshape i\kern-0.25em b}\kern-0.8em\TeX}}}

\setcopyright{iw3c2w3}
\begin{document}
\copyrightyear{2021}
\acmYear{2021} 
\acmConference[WWW '21]{Proceedings of the Web Conference 2021}{April 19--23, 2021}{Ljubljana, Slovenia} 
\acmBooktitle{Proceedings of the Web Conference 2021 (WWW '21), April 19--23, 2021, Ljubljana, Slovenia}
\acmPrice{}
\acmDOI{10.1145/3442381.3450115}
\acmISBN{978-1-4503-8312-7/21/04}

\title{Graph-based Hierarchical Relevance Matching Signals for Ad-hoc Retrieval}

%
\author[Xueli Yu, Weizhi Xu, Zeyu Cui, Liang Wang]{Xueli Yu$^{1,*}$, Weizhi Xu$^{1,2,*}$, Zeyu Cui$^{1,2}$, Shu Wu$^{1,2,\dagger}$, Liang Wang$^{1,2}$}
\makeatletter
\def\authornotetext#1{
	\g@addto@macro\@authornotes{%
	\stepcounter{footnote}\footnotetext{#1}}%
}
\makeatother

\authornotetext{The first two authors contributed equally to this work.}
\authornotetext{Corresponding author.}

\affiliation{%
	\institution{$^1$Center for Research on Intelligent Perception and Computing, Institute of Automation, Chinese Academy of Sciences}
	\institution{$^2$University of Chinese Academy of Sciences}
}

\email{{xueli.yu,weizhi.xu}@cripac.ia.ac.cn,           {zeyu.cui,shu.wu,wangliang}@nlpr.ia.ac.cn}

\def\authors{Xueli Yu, Weizhi Xu, Zeyu Cui, Shu Wu, Liang Wang}
\begin{abstract}
The ad-hoc retrieval task is to rank related documents given a query and a document collection. A series of deep learning based approaches have been proposed to solve such problem and gained lots of attention. However, we argue that they are inherently based on local word sequences, ignoring the subtle long-distance document-level word relationships. To solve the problem, we explicitly model the document-level word relationship through the graph structure, capturing the subtle information via graph neural networks. In addition, due to the complexity and scale of the document collections, it is considerable to explore the different grain-sized hierarchical matching signals at a more general level. Therefore, we propose a \textbf{G}raph-based \textbf{H}ierarchical \textbf{R}elevance \textbf{M}atching model (GHRM) for ad-hoc retrieval, by which we can capture the subtle and general hierarchical matching signals simultaneously. We validate the effects of GHRM over two representative ad-hoc retrieval benchmarks, the comprehensive experiments and results demonstrate its superiority over state-of-the-art methods.
\end{abstract}

\ccsdesc[500]{Information systems~Learning to rank}

\keywords{Hierarchical graph neural networks, Ad-hoc retrieval, Relevance matching}

\maketitle

\section{Introduction}
During recent years, deep learning methods have gained lots of attention in the field of Information Retrieval (IR), where the task is to obtain a list of documents that are relevant to a given query (i.e., a short query and a collection of long documents in the ad-hoc retrieval). Compared to the traditional methods which utilize the hand-crafted features to match between the query and documents, deep learning based approaches can extract the matching patterns between them automatically, thus have been widely applied and have made remarkable success.

Generally speaking, the deep learning based query-document matching approaches can be roughly divided into two categories, i.e., the semantic matching and the relevance matching. In detail, the former focuses on the semantic signals, which learns the embeddings of the query and document respectively, and calculates the similarity scores between them to make the final prediction, such as the methods of DSSM \cite{huang2013learning}, CDSSM \cite{shen2014latent} and ARC-I \cite{hu2014convolutional}. The latter emphasizes more on relevance, like the DRMM \cite{guo2016deep}, KNRM \cite{xiong2017end}, and PACRR \cite{hui2017pacrr,hui2018co}, which capture the relevance signals directly from the word-level similarity matrix of the query and document. As mentioned in \cite{guo2016deep}, the ad-hoc retrieval tasks need more exact signals rather than the semantic one. Therefore, relevance matching methods are more applicable in this scenario. In this paper, we also focus on the relevance matching methods.

\begin{figure}[t]
	\centering
	\includegraphics[scale=0.4]{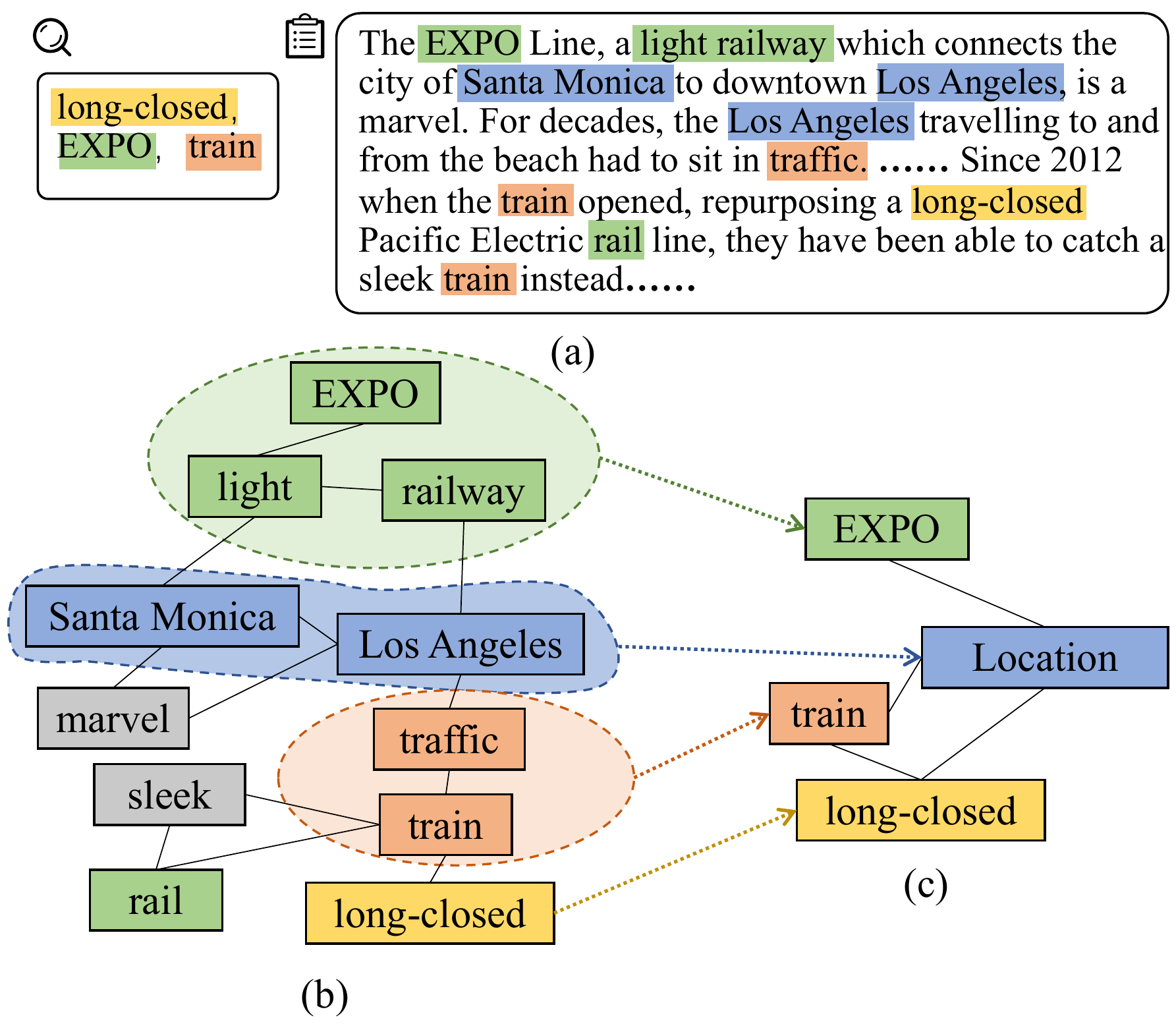}
 	\label{fig:1}
	\caption{An example of the process of hierarchical query-document relevance matching. $\mathbf{(a)}$ The query and the candidate document (some words are omitted). $\mathbf{(b)}$ A graph with a part of words in the document. $\mathbf{(c)}$ A hierarchical graph containing the more critical words and discarding the words unrelated to the query. }
	\label{fig:1}
\end{figure}

While previous relevance matching approaches achieve satisfying results, we argue that the fine-grained long-distance word relationship in the document has not been explored so far. In other words, as terms in the query may not always appear exactly together in the candidate document, it is necessary to consider the subtler document-level word relationship. However, such characteristics are ignored in the existing works which rely on local word sequences \cite{pang2016text,pang2017deeprank,hui2017pacrr}. One solution is to explicitly model the document-level word relationship through the graph structure, such as \cite{yao2019graph, zhang2020every} in the field of text classification, in which the graph neural networks are utilized as a language model to capture the long-distance dependencies. 
Taking Figure \ref{fig:1}$(a)$ and \ref{fig:1}$(b)$ as an example, where $(a)$ presents the query "long-closed EXPO train" and the candidate document, $(b)$ is a graph structure with a part of words in the document. Although the words "long-closed" and "EXPO" are distributed non-consecutively in the document, their relationship can be clearly learned via the high-order connection in the graph structure of $(b)$, which validates the advantage and necessity of utilizing the graph-based methods to capture subtle document-level word relationships.

However, among the above graph-based language models, the different grain-sized hierarchical signals at a more general level are almost ignored, which are also critical characteristics to be considered for ad-hoc retrieval task due to the complexity and scale of the document collections. Taking Figure \ref{fig:1}$(b)$ and \ref{fig:1}$(c)$ as an example, where $(c)$ is the query-aware hierarchical graph extracted from $(b)$. The processing of $(b)$ to $(c)$ could be summarized as two aspects. One is that the words unrelated to the query are dropped, such as "marvel" and "sleek". Another is that the words which may have similar effect for matching the query are integrated to a more critical node. For instance, the words "Santa Monica" and "Los Angeles" in $(b)$ are integrated into the node "Location" in $(c)$, because of their similar location-based supplement to the query. Therefore, considering the different grain-sized interaction information of the hierarchical graphs makes the relevance matching signals more general and comprehensive. Accordingly, to capture the above signals, inspired by the hierarchical graph neural networks methods \cite{ying2018hierarchical, lee2019self}, we introduce a graph pooling mechanism to extract important matching signals hierarchically. Thus, both the subtle and general hierarchical matching signals can be captured simultaneously.  

In this work, we propose a \textbf{G}raph-based \textbf{H}ierarchical  \textbf{R}elevance \textbf{M}atching model (GHRM) to explore different grain-sized query-document interaction information concurrently. 
Firstly, for each query-document pair, we transform the document into the graph-of-words form \cite{rousseau2015text}, in which the nodes indicate words and edges indicate the co-occurrent frequences between each word pairs. For the node feature, we represent it through the interaction between itself and the query term, which can obtain critical interaction matching signals for the relevance matching compared to the traditional raw word feature. Secondly, we utilize the architecture of GHRM to model the different grain-sized hierarchical matching signals on the document graph, where the subtle and general interaction information can be captured simultaneously. Finally, we combine the signals captured in each blocks of the GHRM to obtain the final hierarchical matching signals, feeding them into a dense neural layer to estimate the relevance score.

We conduct empirical studies on two representative ad-hoc retrieval benchmarks, and results demonstrate the effectiveness and rationality of our proposed GHRM\footnote{Code and data available at https://github.com/CRIPAC-DIG/GHRM}. 

In summary, the contributions of this work are listed as follows:
\begin{itemize}
	\item We model the long-distance document-level word relationship via graph-based methods to capture the subtle matching signals.
	\item We propose a novel hierarchical graph-based relevance matching model to learn different grain-sized hierarchical matching signals simultaneously.
	\item We conduct comprehensive experiments to examine the effectiveness of GHRM, where the results demonstrate its superiority over the state-of-the-art methods in the ad-hoc retrieval task.
\end{itemize}

\section{Related Work}
In this section, we briefly review the previous work in the field of ad-hoc retrieval and graph neural networks.

\subsection{Ad-hoc Retrieval}
Ad-hoc retrieval is a task mainly about matching two pieces of text (i.e. a query and a document). The deep learning techniques have been widely utilized in this task, where previous methods can be roughly grouped into two categories: semantic matching and relevance matching approaches. In the former, they propose to embed the representations of query and document into two low-dimension spaces independently, and then calculate the relevance score based on these two representations. For instance, DSSM \cite{huang2013learning} learns the representations via two independent Multi-Layer Perceptrons (MLP) and compute the relevance score as the cosine similarity between the outputs of the last layer of two networks. C-DSSM \cite{shen2014latent} and ARC-I \cite{hu2014convolutional} further captures the positional information by utilizing the Convolutional Neural Network (CNN) instead of MLP. However, this kind of methods are mainly based on the semantic signal, which is less effective for the retrieval task \cite{guo2016deep}. In the latter, relevance matching approaches capture the local interaction signal by modeling query-document pairs jointly. They all follow a general paradigm (i.e. obtaining the interaction signal of the query-document pair first via some similarity functions, and then employing deep learning models to further explore this signal). \citet{guo2016deep} propose DRMM, which utilizes a histogram mapping function and MLP to process the interaction matrix. Later, \cite{hui2017pacrr}, \cite{xiong2017end}, and \cite{hui2018co} employ CNN to capture the higher-gram matching pattern in the document, which considers phrases rather than a single word when interacting with the given query. In addition, a series of multi-level methods \cite{nie2018empirical,rao2019multi} have been proposed, in which \cite{rao2019multi} models word-level CNN and character-level CNN respectively to distinguish the hashtags and the body of the text in the microblog and twitter, capturing multi-perspective information for relevance matching. However, such method is just modeling two different perspectives of the text, not the exact hierarchical matching signals which we will explore in this paper.

Recently, a series of BERT-based methods \cite{macavaney2019cedr, dai2019deeper} have also been proposed in this field, where the BERT’s classification vector is combined with the existing ad-hoc retrieval architectures (using BERT’s token vectors) to obtain the benefits from both approaches.
\subsection{Graph Neural Networks}
Graph neural networks (GNNs) are a promising way to learn the graph representation by aggregating the information from neighborhoods \cite{hu2020graphair}. They can be roughly divided into two lines (i.e. the spectral method \cite{defferrard2016convolutional, kipf2017semi} and the spatial method \cite{velivckovic2018graph, hamilton2017inductive}) regarding different aggregation strategies. Due to the ability that capturing the structural information of data, 
graph neural networks have been widely applied in various domains such as recommendation system \cite{wu2019session, Yu:2020dn, zhang2020personalized,li2019fi} and Natural Language Processing (NLP) \cite{zhang2020every}. They all model the long-distance relationship between items or words via utilizing graph neural networks to explore the graph structure. 
\cite{zhang2018multiresolution} introduces GNNs to the IR task, in which they employ a multi-layer graph convolutional network (GCN) \cite{kipf2017semi} to learn the representations of words in the documents. Nevertheless, it learns the embeddings of the query and document respectively, which is based on the semantic matching rather than relevance matching. 

In recent years, hierarchical graph neural networks, which are proposed to capture signals with different level of the graph through the pooling strategies, have attracted lots of research interest. Several hierarchical pooling approaches have been proposed so far. DiffPool \cite{ying2018hierarchical} learns an assignment matrix that divides nodes into different clusters in each layer, hence reducing the scale of graphs. gPool \cite{gao2019graph} reduces the time complexity by utilizing a projection vector to calculate the importance score of each node. Furthermore, \citet{lee2019self} propose a method namely SAGPool, in which there are three graph convolutional layers followed by three pooling layers. They define the outputs of each pooling layer as attention scores and drop nodes according to the scores. 

In our previous work \cite{zhang2021graph}, we utilize GNNs to model the query-document interaction for ad-hoc retrieval. In this paper, inspired by the progress in the domain of hierarchical graph neural networks, we further design a graph-based hierarchical relevance matching architecture based on graph neural networks and a graph pooling mechanism.

\section{Proposed Method}
\begin{figure*}[t]
\setlength{\abovecaptionskip}{0pt}  
\setlength{\belowcaptionskip}{0pt}  
   \begin{center}
   \includegraphics[width=1\textwidth]{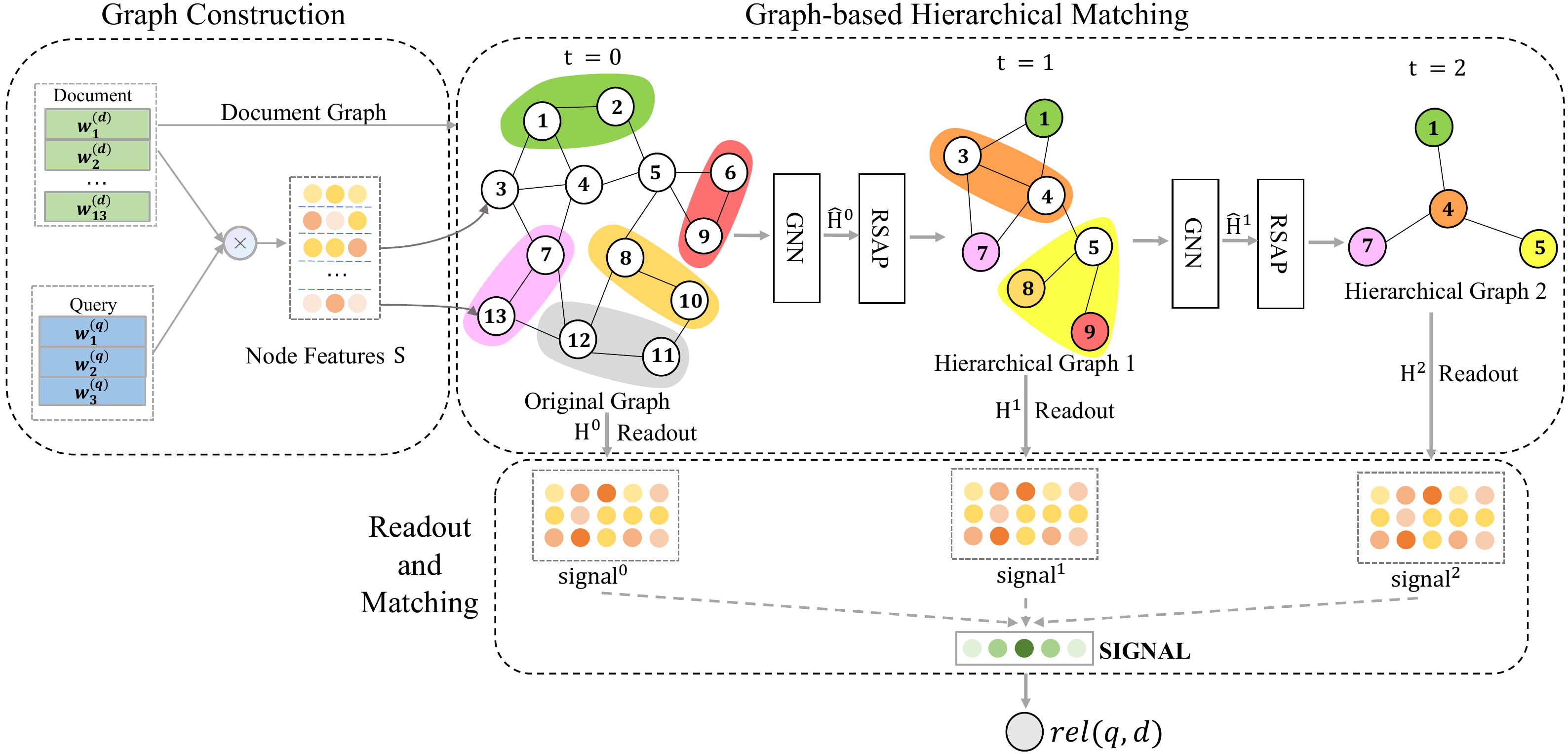}
   \end{center}
   \caption{The architecture of the GHRM model, which consists of the following parts: graph construction, graph-based hierarchical matching, readout and matching and the final relevance score. $\mathbf{(1)}$ \emph{Graph Construction}: The node feature matrix is constructed via the similarity score between the query and the document, in which each node feature represents the interaction signals between its word embedding and the query term embeddings. $\mathbf{(2)}$ \emph{Graph-based Hierarchical Matching}: Through the diffenrent grain-sized hierarchical graphs, the words unrelated to the query are firstly deleted (grey parts of the graph), and several critical nodes which may represent a specific effect on the query (a specific color in the graph) are adaptively preserved via the RSAP mechanism. $\mathbf{(3)}$ \emph{Readout and Matching}: The outputs of the readout layer in each block are combined together to calculate the matching score.} 
   \label{fig:2}
\end{figure*}

In this section, we first give the problem definition and describe how to construct the graph of the document according to its interaction signals with the query. Then, we demonstrate the graph-based hierarchical matching method in detail. Finally, the procedures of model training and matching score prediction are described. Figure \ref{fig:2} illustrates the overall process of our proposed architecture, including the \emph{graph construction}, \emph{graph-based hierarchical matching} and \emph{
readout and matching score prediction}.

\subsection{Problem Formulation}
For each query and document, denoted as $q$ and $d$ respectively, we represent them as a sequence of words $q=\left[w_{1}^{(q)}, \ldots, w_{M}^{(q)}\right]$  and $d=\left[w_{1}^{(d)}, \ldots, w_{N}^{(d)}\right]$, where $w_{i}^{(q)}$ denotes the $i$-th word in the query, $w_{i}^{(d)}$ denotes the $i$-th word in the document, $M$ and $N$ denote the length of the query and the document respectively. In addition, the aim of this problem is to rank a series of relevance scores $rel(q,d)$ regarding the query words and the document words.

\subsection{Graph Construction}
\label{sec:graphconstruct}
To capture the document-level word relationships, we construct a document graph $\mathcal{G}=(\mathcal{V}, \mathcal{E})$, where the $\mathcal{V}$ is the set of vertexes with node features, and $\mathcal{E}$ is the set of edges. The construction procedures of the node feature matrix and the adjacency matrix are described as follows.

\subsubsection{Node feature matrix construction}
In the graph $\mathcal{G}$, each node represents as the word in the document. Hence the word sequence is denoted as a series of node set $\left\{w_{1}^{(d)}, \ldots, w_{n}^{(d)}\right\}$, where $n$ is the number of unique words in the document ($|\mathcal{V}| = n  \leq N$). In addition, to introduce the query-document interaction signals into the graph, we set the node feature as the interaction signals between its word embedding and the query term embeddings, which is similar to \cite{zhang2021graph}. The cosine similarity matrix is applied to represent such interaction matrix, denoted as $\mathbf{S} \in \mathbb{R}^{n \times M}$, where the element ${s}_{ij}$ is set as the similarity score of the node $w^{(d)}_i$ and the query term $w^{(q)}_j$ and it is formulated as:

\begin{equation}{s}_{ij}=cosine\left(\mathbf{e}_i^{(d)}, \mathbf{e}_j^{(q)}\right)
\end{equation}

where $\mathbf{e}_{i}^{(d)}$ and $\mathbf{e}_{j}^{(q)} $ denote the word embedding vectors for $w_{i}^{(d)}$ and $w_{j}^{(q)}$ respectively. Particularly in this work, the word2vec \cite{mikolov2013distributed} method is utilized to convert each word into a dense vector as the initial word embedding.

\subsubsection{Adjacency matrix construction}
As a graph contains nodes and edges, after obtaining the nodes with features, we then focus on the adjacency matrix construction which generally describes the connection and relationships between the nodes. In detail, we apply a sliding window along with the document word sequences $d$, building a bi-directional edge between a word pair if they co-occur within the sliding window. We guarantee that every word can be connected with its neighbor words which may share contextual information via restricting the size of the window. It is worth mentioning that compared to the traditional local relevance matching methods \cite{xiong2017end, hui2017pacrr, hui2018co}, the graph construction method of our GHRM model can further obtain the document-level receptive field by bridging the neighbor words in different hops together. In other words, it can capture the subtle document-level relationship that we concern. 

Formally, the adjacency matrix $\mathbf{A} \in \mathbb{R}^{n \times n}$ is denoted as:
\begin{equation}
\mathbf{A}_{i j}=\left\{\begin{array}{ll}
count(i, j) & \text{if } i \not= j \\
0 & \text{otherwise}
\end{array}\right.
\end{equation}
where $count(i, j)$ is the times that the words $w_{i}^{(d)}$ and $w_{j}^{(d)}$ appear in the same sliding window. 

Furthermore, in order to alleviate the problem of gradient exploding and vanishing, following the study of \cite{kipf2017semi}, the adjacency matrix is normalized as $\tilde{\mathbf{A}} = \mathbf{D}^{-\frac{1}{2}} \mathbf{A} \mathbf{D}^{-\frac{1}{2}}$, where $\mathbf{D} \in \mathbb{R}^{n \times n}$ is the diagonal degree matrix and $\mathbf{D}_{ii} = \sum_j \mathbf{A}_{ij}$.

\subsection{Graph-based Hierarchical Matching}
After the graph $\mathcal{G}$ is constructed, we continue to utilize its node features and structure information with the hierarchical graph neural networks. Specifically, both the subtle and general query-document matching signals are captured mutually following the hierarchical matching structure. As is shown in Figure 2, the architecture of the graph-based hierarchical matching consists of multi-blocks each of which contains a Graph Neural Network (GNN) layer, a Relevance Signal Attention Pooling (RSAP) layer and a readout layer. Through this module, different grain-sized hierarchical matching signals can be captured exhaustively. Finally, the outputs of each block in the graph-based hierarchical matching module are combined together as the hierarchical output. To be specific, we set $\forall t\in [0, T]$ as the $t$-th block of the hierarchical matching, where $T$ is the total number of blocks.

\subsubsection{Graph Neural Network Layer}
We denote the adjacency matrix at $t$-th block as $\mathbf{A}^t \in \mathbb{R}^{m \times m}$, and the node feature matrix at $t$-th block as $\mathbf{H}^{t} \in \mathbb{R}^{m \times M}$, where $m$ is the number of nodes at block $t$ and $M$ is the feature dimension that equals to the number of query terms. As discussed in Section \ref{sec:graphconstruct}, we initialise the $\mathbf{H}^{0}$ with the query-document interaction matrix:
\begin{equation}\mathbf{H}^0 =  \mathbf{S}
\end{equation}
where $\mathbf{H}^{0}_{i}$ denotes the representation of $i$-th node in the graph which equals to $\mathbf{S}_{i}$, i.e., the $i$-th row of the interaction matrix $\mathbf{S}$.

It is crucial for a word to obtain the information from its context since the context is always beneficial for the understand of the center word. In a document graph, one word node can aggregate contextual information from its 1-hop neighborhood, which is formulated as
\begin{equation}\mathbf{a}_{i}^{t}=\sum_{(w_{i}, w_{j}) \in \mathcal{E}} \mathbf{\tilde{A}}^{t}_{ij} \mathbf{W}^{t}_{a} \mathbf{H}_{j}^{t}\end{equation}
where $\mathbf{a}_i^t \in \mathbb{R}^{M}$ denotes the message aggregated from neighbors, $\tilde{\mathbf{A}}^{t}_{i j}$ is the normalized adjacency matrix and $\mathbf{W}^t_a$ is a trainable weight matrix which projects node features into a low-dimension space. The information can be propagated to the $t$-hop neighborhood when we repeat such operation $t$ times. Since the node features are query-document interaction signals, the proposed model can capture the subtle signal interaction between nodes within $t$-hop neighborhood on the document graph via the propagation.


To incorporate the neighborhood information into the word node and also preserve its original features, we employ a GRU-like function \cite{li2016gated}, which can adjust the importance of the current embedding of a node $\mathbf{H}^{t}_i$ and the information propagated from its neighborhoods $\mathbf{a}^{t}_i$, hence its further representation is  $\mathbf{\hat{H}}^{t} = \mathbf{GNN({H}}^{t})$, where the $\mathbf{GNN}$ function is formulated as,
\begin{equation}\begin{array}{l}
\mathbf{z}_{i}^{t}=\sigma\left(\mathbf{W}^{t}_{z} \mathbf{a}_{i}^{t}+\mathbf{U}^{t}_{z} \mathbf{H}_{i}^{t}+\mathbf{b}^{t}_{z}\right)
\end{array}\end{equation}
\begin{equation}
\mathbf{r}_{i}^{t}=\sigma\left(\mathbf{W}^{t}_{r} \mathbf{a}_{i}^{t}+\mathbf{U}^{t}_{r} \mathbf{H}_{i}^{t}+\mathbf{b}^{t}_{r}\right)
\end{equation}
\begin{equation}\tilde{\mathbf{H}}_{i}^{t}=\tanh \left(\mathbf{W}^{t}_{h} \mathbf{a}_{i}^{t}+\mathbf{U}^{t}_{h}\left(\mathbf{r}_{i}^{t} \odot \mathbf{H}_{i}^{t}\right)+\mathbf{b}^{t}_{h}\right)\end{equation}
\begin{equation}\mathbf{\hat{H}}_{i}^{t}=\tilde{\mathbf{H}}_{i}^{t} \odot \mathbf{z}_{i}^{t}+\mathbf{H}_{i}^{t} \odot\left(1-\mathbf{z}_{i}^{t}\right)\end{equation}
where $\sigma(\cdot)$ is the sigmoid function, $\odot$ is the Hardamard product operation, tanh$(\cdot)$ is the non-linear function, and all $\mathbf{W}^{t}_*$, $\mathbf{U}^{t}_*$ and $\mathbf{b}^{t}_*$ are trainable parameters.

In particular, $\mathbf{r}^{t}_i$ represents the reset gate vector, which is element-wisely multiplied by the hidden state $\tilde{\mathbf{H}}^{t}_i$ to generate the information to be forgot. Besides, $\mathbf{z}^{t}_i$ determines which component of the current embeddings to be pushed into next iteration. Notably, we have also tried another message passing model, i.e., GCN \cite{kipf2017semi} in our experiments but did not observe satisfying performance.

\subsubsection{Relevance Signal Attention Pooling Layer}
As attention mechanisms are widely used in the field of deep learning \cite{velivckovic2018graph, vaswani2017attention}, which makes it possible to focus more on the important features than the relatively unimportant ones. Therefore, it is considerable to utilize such mechanism to the graph pooling layer, by which the important graph nodes and different grain-sized interaction matching signals can be explored. Inspired by previous hierarchical GNN methods \cite{lee2019self,ying2018hierarchical,gao2019graph}, we introduce a Relevance Signal Attention Pooling mechanism (RSAP) into the pooling layer, obtaining the attention scores of each node via the graph neural network. As shown in Figure \ref{fig:2}, through the RSAP, the hierarchical graph in $t=1$ and hierarchical graph in $t=2$ can discard the words which are unrelated to the query (like the grey nodes in the original graph), and adaptively preserve the critical nodes that can represent a specific effect on the query. In detail, the attention score $\mathbf{P}^{t} \in \mathbb{R}^{m}$ denoting the attention score of $m$ nodes in $t$-th block is calculated as follows,

\begin{equation}
\mathbf{P}^{t}=\mathbf{GNN}\mathbf{(\hat{H}}^{t} \mathbf{\cdot {W}}^{t}_{p})
\end{equation}
where $\mathbf{GNN}$ is the graph neural network function the same as mentioned above, the $\mathbf{W}^{t}_{p}$ is an trainable attention matrix.

Once the attention scores of each node are obtained, we then focus on the important node selection via the hard-attention mechanism. Following the method of \cite{lee2019self,gao2019graph,cangea2018towards}, we retain a portion of nodes in the document graph, which represent critical signals in a more general level. Through this hard-attention mechanism, the words which are unrelated to the query are filtered out. The pooling ratio ${rate} \in\mathbf(0,1]$ is a hyperparameter, which determines the number of nodes to keep in each RSAP layer. The top $\lceil m \cdot rate \rceil$ \footnote{$\lceil \cdot \rceil$ denotes the round-up operation (e.g., $\lceil 3.3 \rceil$ = $4$).} nodes are selected based on the value of $\mathbf{P}^{t}$. 

\begin{equation}
idx = \mathbf{top \_ rank}({ \mathbf{P}^{t},\lceil m \cdot rate \rceil)}
\end{equation}
\begin{equation}
\mathbf{P}^{t}_{mask} = \mathbf{P}^{t}_{idx}
\end{equation}
\begin{equation}
\mathbf{A}^{t+1} = \mathbf{A}^{t}_{idx,idx}, \quad \mathbf{H}^{'^t} = \mathbf{\hat{H}}_{idx}^{t}
\end{equation}
where $\mathbf{top} \_ \mathbf{rank}(\cdot)$ is the function that returns the indices of the top $\lceil m \cdot rate \rceil$ values, ${\cdot}_{idx}$ is an indexing operation and $\mathbf{P}^{t}_{mask}$ is the attention mask, where elements are set to 0 if the nodes are discarded according to the $\mathbf{top} \_ \mathbf{rank}(\cdot)$ operation. $\mathbf{A}^{t}_{idx,idx}$ is the row-wise and column-wise indexed adjacency matrix. $\mathbf{\hat{H}}^{t}_{idx}$ is the row-wise (i.e., node-wise) indexed feature matrix from $\mathbf{\hat{H}}^t$.

Next, the soft-attention mechanism is applied on the pooling operation based on the selected important nodes, and the new feature matrix $\mathbf{H}^{t+1}$ which is fed into the the $(t+1)$-th block is calculated as follows,

\begin{equation}
\mathbf{H}^{t+1} = \mathbf{H}^{'^t} \odot \mathbf{P}^{t}_{mask}
\end{equation}
where $\mathbf{\odot}$ is the broadcasted element-wise product, realizing the soft-attention operation, through which the critical query-document interaction matching signals are further emphasized. 


\subsubsection{Readout Layer}
In order to aggregate the node features to make a fixed size representation as the query-document relevance signal, we select a fix-sized number of features from ${H}^t$ in each block through the method of $k$-max-pooling strategy on the dimension of query. The formulas are written as follows,

\begin{equation} \label{equ:top}
\mathbf{signal}^{t} = \mathbf{topk}(\mathbf{H}^{t})
\end{equation}
where 
the function of $\mathbf{topk(\cdot)}$ is operated on the column-wise dimension of $\mathbf{H}^{t}$, denoting the top $k$ values for each term of the query respectively. Hence $\mathbf{signal}^{t}$ is the output relevance matching signal of $t$-th block.

After obtaining the fix-sized query-document relevance signal of each block, we then combine each block's relevance matrix together as the hierarchical relevance matching signals, which is formulated as,

\begin{equation}
\mathbf{SIGNAL}=\mathbf{signal}^{0} \parallel \mathbf{signal}^{1} \parallel \ldots \parallel \mathbf{signal}^{t} \parallel \ldots \parallel \mathbf{signal}^{T
}\end{equation}

where $\mathbf{SIGNAL}\in \mathbb{R}^{k(T+1)\times M}$ represents the overall hierarchical relevance signal, which can capture both the subtle and the general query-document matching signals simultaneously. Specially, the $\mathbf{signal}^{0}$ denotes the $\mathbf{topk}(\mathbf{H}^{0})$ from the initial similarity matrix of the query and document.

\subsection{Matching Score and Model Training}
To convert the hierarchical relevance signals into the actual relevance scores for training and inference, we input the relevance matrix $\mathbf{SIGNAL}$ into the further deep neural networks. Since the elements in each column of $\mathbf{SIGNAL}$ are the relevance signals of each corresponding query word, considering that different query words may have different importances for retrieval, we assign the relevance signals corresponding to each query word with a soft gating network \cite{guo2016deep} as, 
\begin{equation}g_{j}=\frac{\exp \left({c} \cdot idf_j \right)}{\sum_{i=1}^{M} \exp \left({c} \cdot idf_i \right)}\end{equation}
where $g_j$ is the corresponding term weight, $idf_j$ is the inverse document frequency of the $j$-th query term, and $c$ is a trainable parameter. Furthermore, we score each term of the query with a weight-shared MLP to reduce the parameters amount and avoid over-fitting, summing the results of it up as the final result, 
\begin{equation}{rel}(q, d)=\sum_{j=1}^{M} g_j \cdot f(\mathbf{SIGNAL}_{j})\end{equation}
where $f(\cdot)$ is a MLP in our model.

Finally, the pairwise hinge loss is adopted for training and optimizing the model parameters, which is widely used in the field of information retrieval, formulated as,
\begin{small}
	\begin{equation}\mathcal{L}\left(q, d^{+}, d^{-}\right)=\max \left(0, 1-rel\left(q, d^{+}\right)+rel\left(q, d^{-}\right)\right)\end{equation}
\end{small}
where $\mathcal{L}\left(q, d^{+}, d^{-}\right)$ is the pairwise loss based on a triplet of the query $q$, a relevant (positive) document sample $d^+$, and an irrelevant (negative) document sample $d^-$.

\section{EXPERIMENTS}
In this section, we conduct experiments on two ad-hoc datasets to answer the following questions:
\begin{itemize}
	\item RQ1: How does GHRM perform compared with the previous relevance matching baselines? 
	\item RQ2: How does the different grain-sized hierarchical signals affect the performance of the model?
	\item RQ3: How does GHRM perform under different hyperparameter settings?
\end{itemize}

\subsection{Experiment Setup}
\subsubsection{Datasets.}
In this part, we briefly introduce two datasets used in our experiments, named Robust04 and ClueWeb09-B.
\begin{itemize}
    \item Robust04\footnote{https://trec.nist.gov/data/cd45/index.html}. There are 250 queries and 0.47M documents, which are from TREC disk 4 and 5 in this dataset.
    \item ClueWeb09-B\footnote{https://lemurproject.org/clueweb09/} is one of the subset from the full data collection ClueWeb09. There are 50M documents collected from web pages and 200 queries. The topics of these texts are obtained from TREC Web Tracks 2009-2012.
\end{itemize}
In both datasets, the train data consists of several query-document pairs, where a query has a most related document (i.e., the ground-truth label). In addition, there are two parts in a query, i.e., a short keyword title and a longer text description, and we only utilize the title in our experiments. Table \ref{tab:1} summarises the statistic of the two datasets.

\subsubsection{Compared Methods.}
To evaluate the performance of our proposed model GHRM, we compare it with a variety of baselines, including traditional language models (i.e., query likelihood model and BM25), deep relevance matching models (i.e., MatchPyramid, DRMM, KNRM, PACRR and Co-PACRR) and a pre-trained BERT-based method (i.e., BERT-MaxP). The brief introduction of each baseline model is presented as follows, 
\begin{itemize}
    \item \textbf{QLM} (Query likelihood model) \cite{zhai2004study} is based on Dirichlet smoothing and have achieved convincing results in the domain of NLP when the deep learning technique has not appeared yet.
    \item \textbf{BM25} \cite{robertson1994some} is a famous and effective bag-of-words model, which is based on and the probabilistic retrieval framework.
    \item \textbf{Pyramid} (MatchPyramid) \cite{pang2016text} first builds up the interaction matrix between a query and a document, then they employ CNN to process the matrix, extracting the different orders of matching features. 
    \item \textbf{DRMM} \cite{guo2016deep} is the pioneer work of the relevance matching approaches. They perform a histogram pooling over the local query-document interaction matrices to summarize the different relevance features. 
    \item \textbf{KNRM} \cite{xiong2017end} applies a kind of kernel pooling to explore the matching features existing in the interaction matrix.
    \item \textbf{PACRR} \cite{hui2017pacrr} redesigns CNNs in terms of kernel size and convolution direction to make CNNs more suitable for the IR task. This model finally utilizes a RNN to capture the long-term dependency over different signals.
    \item \textbf{Co-PACRR} \cite{hui2018co} is a variant of PACRR, which takes the contextual matching signals into account, and achieves a better result than PACRR.
    \item \textbf{BERT-MaxP} \cite{dai2019deeper} utilizes BERT to deeply understand the text for the relevance matching task, demonstrating that the contextual text representations from BERT are more effective than traditional word embeddings. 
\end{itemize}

\begin{table}[]
	\footnotesize
	\begin{tabular}{@{}ccccc@{}}
		\toprule
		\textbf{Dataset}     & \textbf{Genre} & \textbf{\# Queries} & \textbf{\# Documents} & \textbf{Avg.length} \\ \midrule
		\textbf{Robust04}    & news           & 250                 & 0.47M                & 460                         \\
		\textbf{ClueWeb09-B} & webpages       & 200                 & 50M                 & 1506                        \\ \bottomrule
	\end{tabular}
	\caption{Statistics of datasets. \# means the number of. M means a million and Avg.length means the average length of documents.}
	\label{tab:1}
\end{table}

\subsubsection{Implementation Details.}
For text preprocessing, by using the WordNet\footnote{https://www.nltk.org/howto/wordnet.html} toolkit, we first make all words in the document and query in the two datasets white-space tokenised, lowercased, and lemmatised. Secondly, we discard the words appear less than ten times in the corpus, which is a normal preprocessing operation in NLP tasks. Following the previous work \cite{hui2017pacrr}\cite{macavaney2019cedr}, we truncate the first 300 and 500 words in each document, the first 4 and 5 words in each query for Robust04 and CluwWeb09-B respectively. The lengths are different for two datasets simply because texts in ClueWeb09-B are almost longer than those in Robust04 and may have more useful information. We utilize the zero-padding if the length of a document or a query is less than the truncated length. In addition, we initialize the word embeddings as the output 300-dimension vectors of the Continuous Bag-of-Words (CBOW) model \cite{mikolov2013distributed} on both two datasets. Also, except for those models that do not need word embeddings, we use the same initialized embeddings to keep the fair comparison. 
All baseline models are implemented, closely following all settings reported in their original paper.

Following the setting in the previous work \cite{macavaney2019cedr}, we divide the both two datasets into three parts: sixty percents of data for training, twenty percents of data for validation and the rest of data for testing. Based on this distribution, we randomly divide the datasets for five times to generate five folds with different data splits (i.e., the training data is different in each fold). Then we utilize the data in a round-robin fashion as \citet{macavaney2019cedr} does. Eventually, we take the average result of five folds as the final performance of the model.

There are a series of hyperparameters in our model including the number of blocks $T$ in the graph-based hierarchical matching module, the pooling ratio $rate$ of $\mathbf{top\_rank}(\cdot)$ in the RASP layer, the $k$ value of $\mathbf{topk}(\cdot)$ in the readout layer, the learning rate and the batch size. They are all tuned on the validation set using the grid search algorithm. In the based model GHRM, we set the number of blocks $T$ as 2, the pooling ratio $rate$ as 0.8, and the number of values $k = 40$. We train the model with a learning rate of 0.001 using the Adam optimizer \cite{adam2015} for 300 epochs. In each epoch, there are 32 batches and each batch contains 16 positive sampled pairs and 16 negative pairs. We rerank the top 150 candidates generated by BM25 Anserini toolkit\footnote{https://github.com/castorini/anserini} on the stage of testing, which is a normal way to test the model in the IR task.

All experiments are conducted using PyTorch 1.5.1 on a Linux server equipped with 4 NVIDIA Tesla V100S GPUs (with 32GB memory each) and 12 Intel Xeon Silver 4214 CPUs (@2.20GHz).

\subsubsection{Evaluation Methodology.}
We utilize two evaluation matrices in our experiments. One is the normalised discounted cumulative gain at rank 20 (nDCG@20) and another is the precision at rank 20 (P@20). Both of them are often used in this kind of ranking task.

\subsection{Model Comparison (RQ1)}
The performance of each model on two datasets is clearly shown in Table \ref{tab:modelcmp}. Based on these results, we have some observations as follows:
\begin{itemize}
	\item First of all, GHRM outperforms both traditional language models and deep relevance matching models by a significant margin. To be specific, compared to the strong baseline Co-PACRR, GHRM advances the performance of nDCG@20 and P@20 by 5.6\% and 2.9\% respectively on Robust04. Besides, on ClueWeb09-B, it achieves an improvement of 15.1\% on nDCG@20 and 10.8\% on P@20, compared to another convincing model DRMM. There are two reasons that may contribute to the improvement: on the one hand, the applying of the graph neural networks can capture the subtle document-level word relationship via extracting all non-consecutively distributed relevant information, yet previous CNN-based models can not capture them. On the other hand, the model with a hierarchical architecture is able to attentively discard some useless information from noisy neighbors and preserve the most important information, thus capturing the different grain-sized hierarchical relevance matching signals. Owing to such two advantages, the relevance matching signals can be obtained comprehensively and the performance of the model is enhanced. 
	
	\item Compared to BERT-MaxP, the results show that even GHRM does not depend on the pre-trained word embeddings, it also gains the comparative performance. In detail, GHRM performs better than BERT-MaxP on ClueWeb09-B while worse on Robust04. The reason may be that there are several characteristic differences between the two datasets. On the one hand, the language style of Robust04 is more formal, making the pre-trained word embeddings of BERT-MaxP more superior. On the other hand, the length of the documents in ClueWeb09-B is relatively long, which may weaken the performance of BERT-MaxP since it restricts the input sequence length as a maximum of only 512 tokens. Meanwhile, the GHRM's advantage of capturing the subtle long-distance word relationships can be represented on  ClueWeb09-B.
	
	\item We also observe that the performance of local relevance matching models slightly fluctuate around the performance of BM25, except the models DRMM and KNRM on ClueWeb09-B. This may mainly because DRMM and KNRM utilize global pooling strategies while others only focus on local relationship. It further validates that only considering the local interaction is insufficient for the ad-hoc retrieval task, the more exhaustive information contained in the different grain-sized hierarchical matching signals may also play a central role.
	
	\item Another observation is that traditional approaches QL and BM25 still outperform some deep learning methods, which demonstrates that the exact matching signal is significant for the ad-hoc retrieval, which has been pointed out by \citet{guo2016deep}. That is why we preserve the initial similarity matrix of the query and document as the first block of matching signal in GHRM. Besides, traditional models also avoid overfitting the train data. 
\end{itemize}                       

\label{sec:modelcompare}
\begin{table}[]
	\fontsize{9.3pt}{11pt}\selectfont
    \begin{tabular}{@{}cllll@{}}
    \toprule
    \multirow{2}{*}{Model} & \multicolumn{2}{c}{Robust04}                           & \multicolumn{2}{c}{ClueWeb09-B}                        \\ \cmidrule(l){2-5} 
                           & \multicolumn{1}{c}{nDCG@20} & \multicolumn{1}{c}{P@20} & \multicolumn{1}{c}{nDCG@20} & \multicolumn{1}{c}{P@20} \\ \midrule
    QL                     & 0.415$^-$                   & 0.369$^-$                & 0.224$^-$                   & 0.328$^-$                \\
    BM25                   & 0.418$^-$                   & 0.370$^-$                & 0.225$^-$                   & 0.326$^-$                \\ \midrule
    MP                     & 0.318$^-$                   & 0.278$^-$                & 0.227$^-$                   & 0.262$^-$                \\
    DRMM                   & 0.406$^-$                   & 0.350$^-$                & 0.271$^-$                   & 0.324$^-$                \\
    KNRM                   & 0.415$^-$                   & 0.359$^-$                & 0.270$^-$                   & 0.330$^-$                \\
    PACRR                  & 0.415$^-$                   & 0.371$^-$                & 0.245$^-$                   & 0.278$^-$                \\
    Co-PACRR               & 0.426$^-$                   & 0.378$^-$                & 0.252$^-$                   & 0.289$^-$
    \\ \midrule
    BERT-MaxP & \textbf{0.469} & - & 0.293 & -
    \\ \midrule
    GHRM                   & 0.450                        & \textbf{0.389}                    & \textbf{0.312}                       & \textbf{0.359}                    \\ \bottomrule
    \end{tabular}
	\caption{The performance of our proposed model GHRM and baselines. The highest performance on each dataset and metric is highlighted in boldface. Significant performance degradation with respect to GHRM is indicated (-) with p-value $\leq$ 0.05.}
	\label{tab:modelcmp}
\end{table}

\subsection{Study of Hierarchical Signals (RQ2)}
\label{sec:graphstructure}
To prove the effectiveness of hierarchical signals in the ad-hoc retrieval task, we further conduct a comparison experiment to study what effects do the different grain-sized hierarchical signals take to the model. In detail, we discard all pooling layers (i.e., RSAP layers) in GHRM, so that all the words in the document are considered equally important along the multi-layer graph neural networks. In addition, we ensure that the graph structure is fixed during the whole training process and no hierarchical signal is generated. We denote this model as GHRM-nopool. For a fair comparison, we keep all other settings the same in GHRM and GHRM-nopool. 

\label{sec:modelcompare}
\begin{table}[]
	\fontsize{9.3pt}{11pt}\selectfont
    \begin{tabular}{@{}cllll@{}}
    \toprule
    \multirow{2}{*}{Model} & \multicolumn{2}{c}{Robust04}                           & \multicolumn{2}{c}{ClueWeb09-B}                        \\ \cmidrule(l){2-5} 
                           & \multicolumn{1}{c}{nDCG@20} & \multicolumn{1}{c}{P@20} & \multicolumn{1}{c}{nDCG@20} & \multicolumn{1}{c}{P@20} \\ \midrule
    GHRM-nopool               & 0.437                   & 0.378                & 0.300                   & 0.351              \\
    GHRM                   & 0.450                        & 0.389                    & 0.312                       & 0.359                     \\ \midrule
    Improv.        & 2.97\%   & 2.91\%    & 4.00\%  & 2.28\%    \\ \bottomrule
    \end{tabular}
	\caption{The comparison of performance between GHRM and GHRM-nopool on Robust04 and ClueWeb09-B. The improvement in terms of percentage is shown in the last row.}
	\label{tab:hierarchical study}
\end{table}

As illustrated in Table \ref{tab:hierarchical study}, it is apparently seen that GHRM outperforms GHRM-nopool by a significant margin on the two datasets and evaluation matrices. Specifically, on Robust04, GHRM outperforms the non-hierarchical model (i.e., GHRM-nopool) by 2.97\% and 2.91\% on the metric of nDCG@20 and P@20 respectively. On ClueWeb09-B, GHRM improves the performance by 4\% on nDCG@20 and 2.28\% on P@20. This reveals that the different grain-sized hierarchical matching signals obtained via GHRM are also critical for retrieving relevant documents. It is worth mentioning that even without the hierarchical signals, the GHRM-nopool still outperforms the traditional language models and deep relevance matching models substantially, which demonstrates the superiority of the document-level word relationship over the local-based relevance matching methods. In addition, based on these subtle information, the various grain-sized hierarchical signals can play as a strong supplement in a more general level, hence improving the performance of relevance matching further.

\subsection{Ablation Study (RQ3)}
\label{sec:neighbouraggre}
In this section, we discuss about the specific hyperparameter setting in GHRM, including the pooling ratio $rate$, the number of blocks $T$ and the number of $k$ in the readout layer.  
\subsubsection{The pooling ratio $rate$}
\begin{figure}[t]
  \centering
  \subfloat{
    \includegraphics[width=0.23\textwidth]{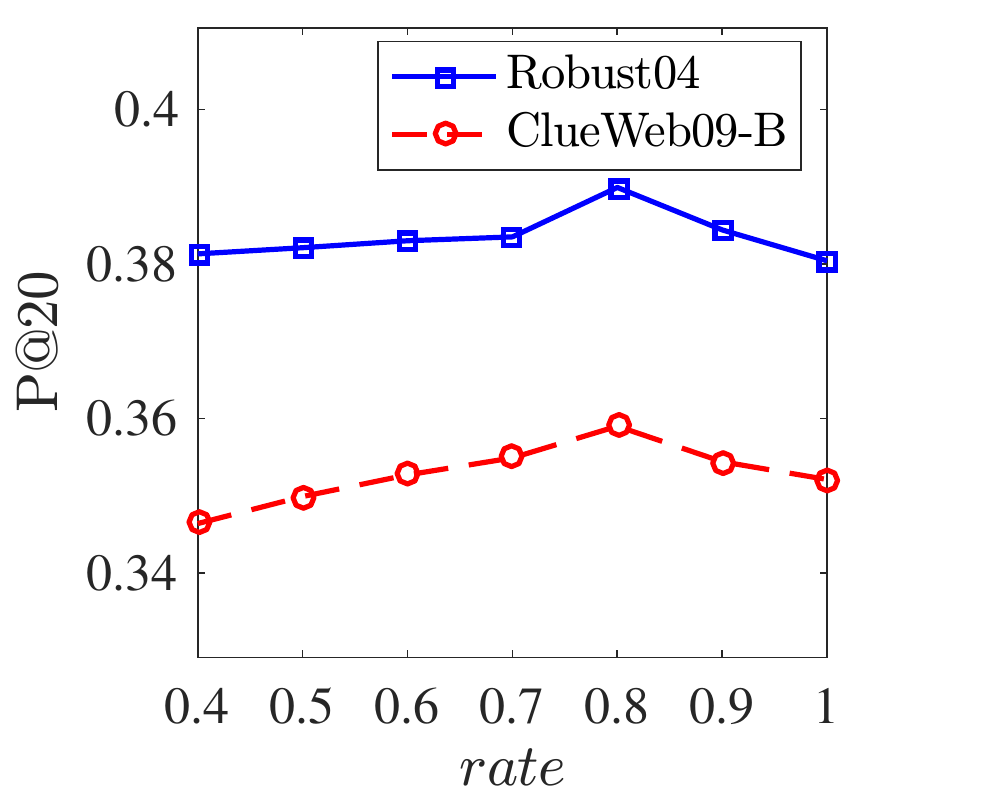}}
  \subfloat{
    \includegraphics[width=0.23\textwidth]{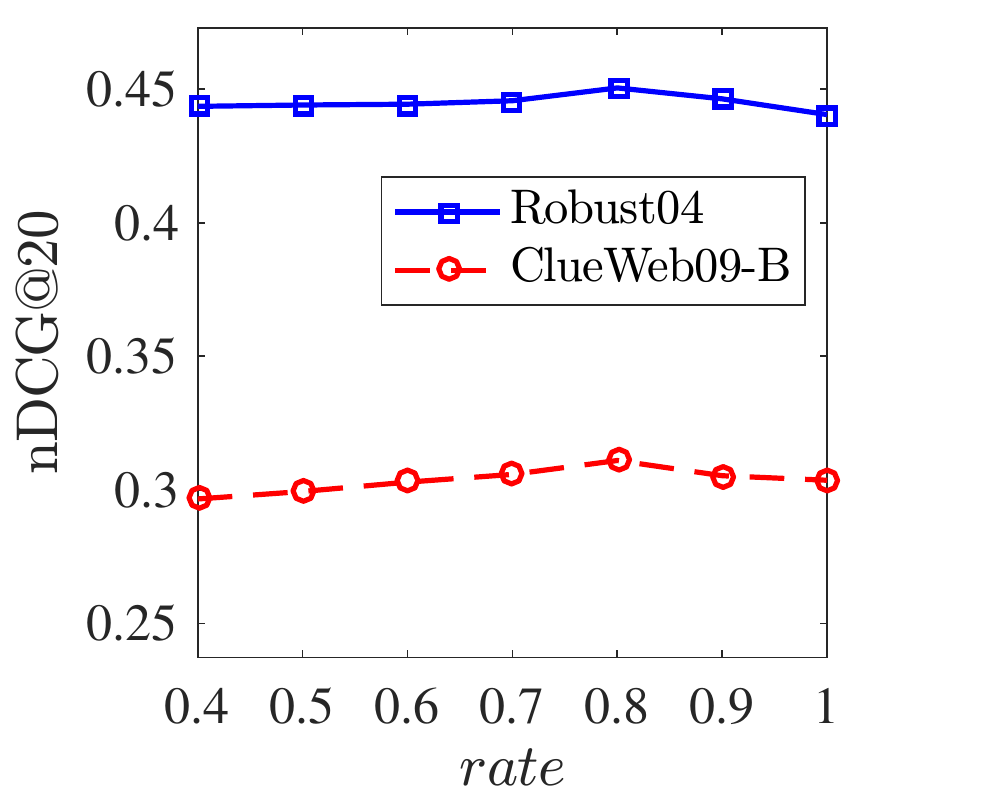}}
  \caption{Influence of different pooling ratios. The model peaks at the best results on both datasets and evaluation matrices when $rate=0.8$ .}
  \label{fig:poolingrate}
\end{figure}

It is an important hyperparameter in our proposed model since it controls the number of critical nodes selected in each RSAP layer. For example, if the $rate = 0.4$, we select 40\% of nodes and discard rest of nodes in the RSAP layer in each block. As is shown in Figure \ref{fig:poolingrate}, the performance first grows up continuously when the $rate$ ranges from 0.4 to 0.8 and then decreases slightly when the $rate$ is over 0.8 on both two datasets. The observations could be listed as follows:
\begin{itemize}
    \item GHRM with a pooling ratio $rate$ of 0.8 peaks at the highest result on both two datasets, which could be due to the suitable amount of deleted nodes. It means that using this rate, we could obtain suitable hierarchical signals from the RSAP layers since those nodes who contribute relatively little to the model training would be deleted.
    \item A low pooling ratio $rate$ may not advance the performance of the model. For example, the model with $rate = 0.4$ has the worst performance of 0.297 and 0.381 on nDCG@20 on ClueWeb09-B and Robust04 respectively. It is probably because that some valuable nodes are deleted and the graph topology becomes sparse, preventing the model from capturing the long-distance word relationship.
    \item When the $rate$ equals to 1.0, we denote the model as GHRM-soft and it can be regarded as matching the query and document only with the soft-attention mechanism since nodes are not discarded in each layer. It is worth noting that GHRM-soft is not the same as GHRM-nopool. The difference is that each signal from nodes are equally processed in GHRM-nopool while soft-attention scores are applied to distinguish each node in GHRM-soft. The performance of the two models implies that different signals should be considered attentively before combining them to output the relevance score. 
    \item The overall results illustrate that the hierarchical matching signals obtained by designing proper pooling ratio $rate$ are important for the ad-hoc retrieval. With proper pooling ratio, the GHRM can mutually capture both subtle and general interaction information between the query and the document, making the matching signals more exhaustive. 
\end{itemize}

\begin{figure}[t]
    \centering
    \subfloat{
    \includegraphics[width=0.23\textwidth]{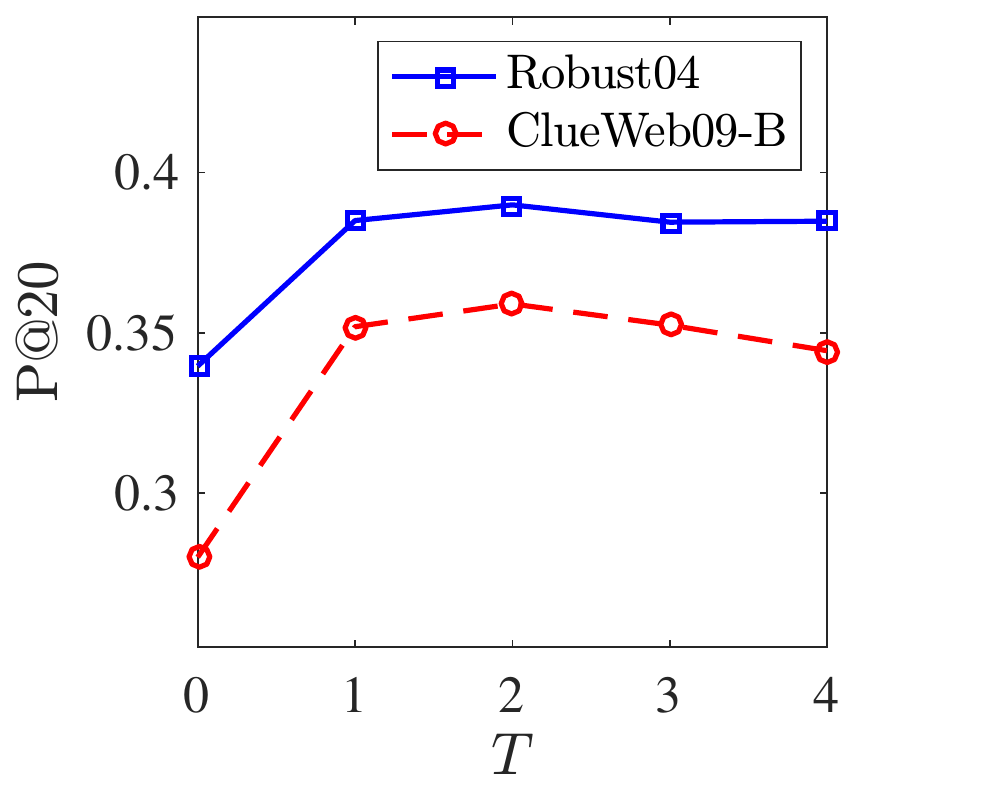}}
    \subfloat{
    \includegraphics[width=0.23\textwidth]{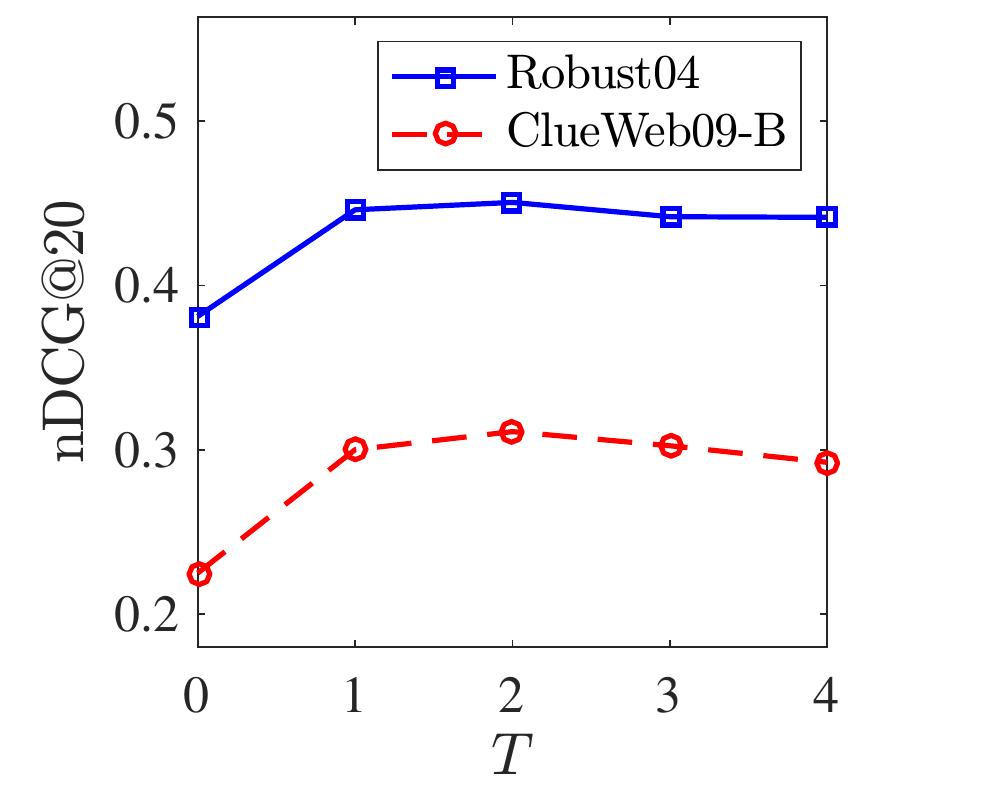}}
    \caption{Influence of the number of blocks in the graph-based hierarchical matching module. The best performance is achieved when $T=2$ on both datasets and evaluation matrices.}
    \label{fig:layernum}
\end{figure}

\subsubsection{The number of blocks $T$ in the graph-based hierarchical matching module}
The number of blocks $T$ is also a critical hyperparameter in GHRM, which decides the extent of different grain sizes learned in the hierarchical matching signals. In this part, we perform experiments on the GHRM from $T=0$ to $T=4$ blocks respectively as shown in Figure \ref{fig:layernum}. We have some observations as follows:
\begin{itemize}
    \item An improvement can be seen from $T=0$ to $T=1$ in Figure \ref{fig:layernum}. When $T=0$, the matching signal represents the one obtained from the initial similarity matrix of the query-document pair. It reveals that the long-distance information in document-level word relationships, which are captured via the graph neural network
    are significant for the query-document matching.
    \item 
    The performance of model grows with the increasing from $T=0$ to $T=2$, which further illustrates the positive effect that different grain-sized hierarchical matching signals take to the model.
    \item We can also see that the performance decreases, when $T$ is over 2. The reason could be that nodes may receive noisy information from high-order neighbors which deteriorates the performance of the model when the number of blocks continue to grow. The 2-hop neighborhood information is sufficient to capture the most significant part of word relationships.
\end{itemize}

\subsubsection{The number of  top $k$ values in the readout layer}
\begin{figure}[t]
  \centering
  \subfloat{
    \includegraphics[width=0.23\textwidth]{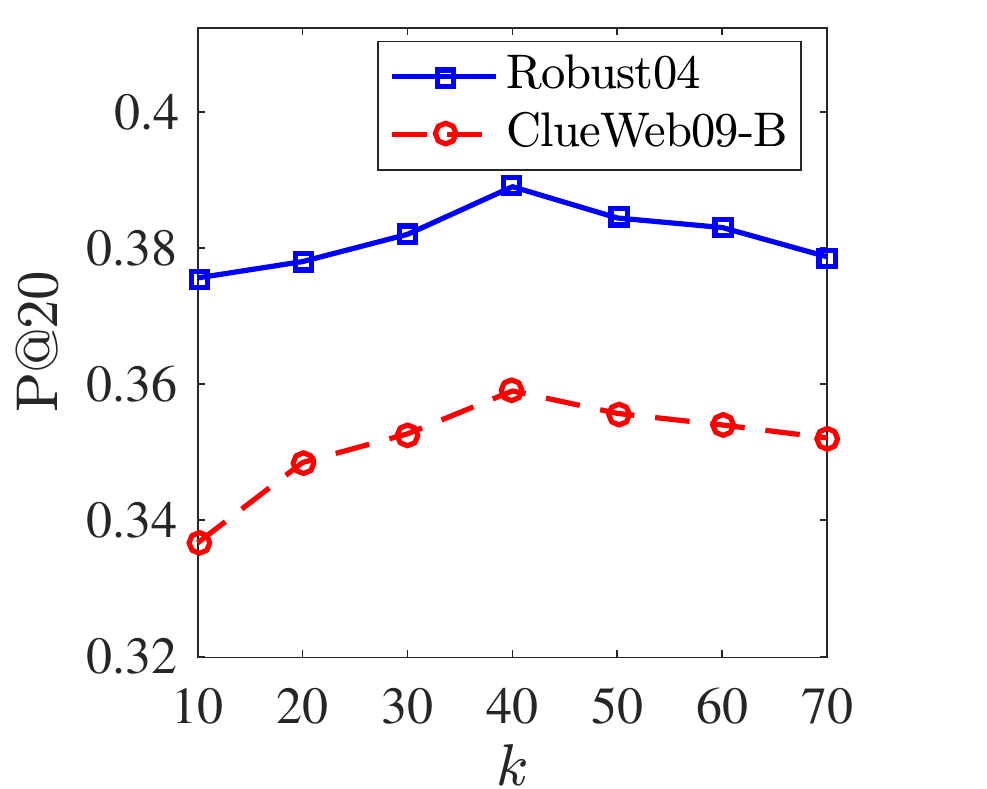}}
  \subfloat{
    \includegraphics[width=0.23\textwidth]{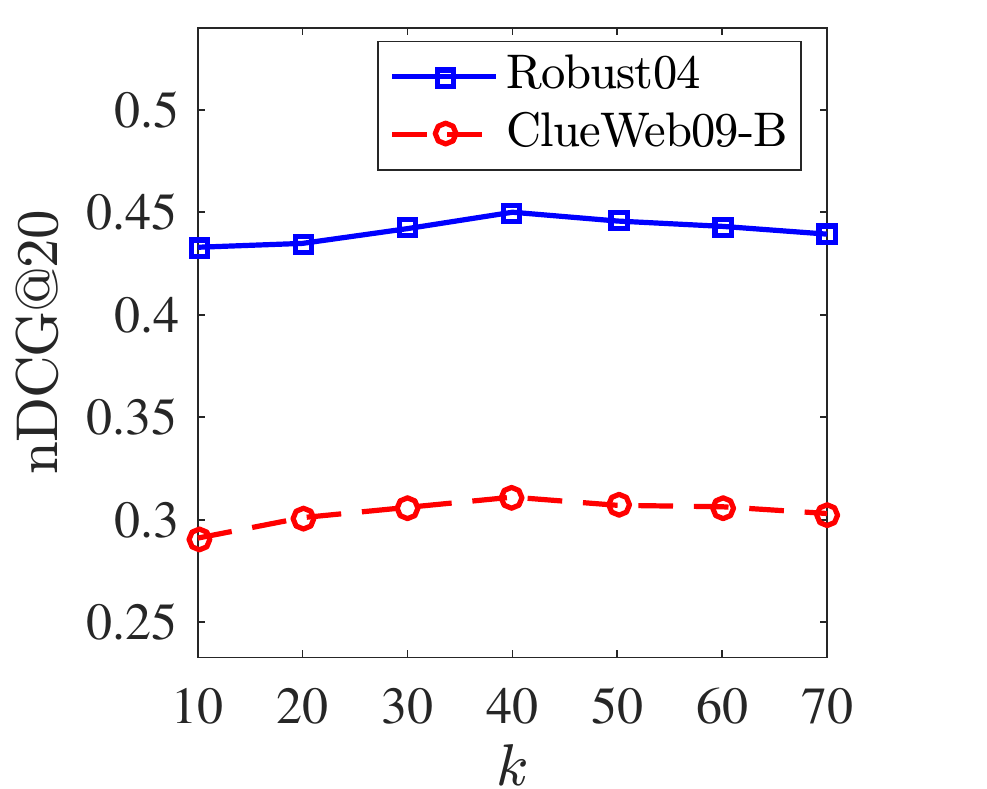}}
  \caption{Influence of the number of  top $k$ values in the readout layer. The best performance is achieved when $k=40$ on both datasets and evaluation matrices.}
  \label{fig:kvalue}
\end{figure}

We also explore the effect of the size of the features that are output from the readout layer of each block in the hierarchical matching module. Figure \ref{fig:kvalue} summarises the experimental performance in terms of different $k$ values of $\mathbf{topk(\cdot)}$ in Equation \ref{equ:top}. From the figure, we have the following observations: 
\begin{itemize}
	\item There is a moderate growth of performance when $k$ is ranged from 10 to 40, which implies that some important hierarchical matching signals are wrongly discarded when the $k$ value is small. Furthermore, when continually enlarging the $k$ value, the GHRM could distinguish more relevant hierarchical matching signals from the relatively irrelevant one.
	\item The performance begins to decline when $k$ continues to grow, which demonstrates that the large size of the readout features may bring some noisy information, such as the bias influence of the document length.
    
	\item It is worth noting that almost all model variants of GHRM with different $k$ values (except $k=10$) exceed the baselines in Table \ref{tab:modelcmp}. This implies that different grain-sized graph-based hierarchical signals are effective for correctly matching the query and document.
\end{itemize}

\section{Conclusion}
In this paper, we introduce a graph-based hierarchical relevance matching method for ad-hoc retrieval named GHRM. By utilizing the hierarchical graph neural networks to model different grain-sized matching signals, we can exactly capture the subtle and general hierarchical interaction matching signals mutually. Extensive experiments on the two representative ad-hoc retrieval benchmarks demonstrate the effectiveness of GHRM over various baselines, which validates the advantages of applying graph-based hierarchical matching signals to ad-hoc retrieval.

\section{Acknowledgments}
This work is supported by National Key Research and Development Program (2018YFB1402605, 2018YFB1402600), National Natural Science Foundation of China (U19B2038, 61772528), Beijing National Natural Science Foundation (4182066).

\bibliographystyle{ACM-Reference-Format}
\bibliography{www.bib}

\end{document}